\documentclass[floatfix,twocolumn,showpacs,preprintnumbers,amsmath,amssymb,prl,superscriptaddress,longbibliography]{revtex4-1}
\usepackage{color}
\usepackage[usenames,dvipsnames,svgnames,table]{xcolor}
\usepackage[colorlinks=true,linkcolor=blue,urlcolor=blue,citecolor=blue]{hyperref}

\usepackage{mathtools}
\usepackage{graphicx}
\usepackage{dcolumn}
\usepackage{array}
\usepackage{lipsum}
\usepackage{bm}
\usepackage{subfigure}
\usepackage{amssymb}
\usepackage{multirow}
\usepackage{tabularx}
\usepackage{amsmath}
\usepackage{braket}
\graphicspath{{plots/}}

\begin{document}
\title{
Effective Static Approximation: A Fast and Reliable Tool\\ for Warm Dense Matter Theory
}

\author{Tobias Dornheim}
\email{t.dornheim@hzdr.de}

\affiliation{Center for Advanced Systems Understanding (CASUS), D-02826 G\"orlitz, Germany}

\author{Attila Cangi}

\affiliation{Center for Advanced Systems Understanding (CASUS), D-02826 G\"orlitz, Germany}

\author{Kushal Ramakrishna}

\affiliation{Helmholtz-Zentrum Dresden-Rossendorf (HZDR), D-01328 Dresden, Germany}

\affiliation{Technische  Universit\"at  Dresden,  D-01062  Dresden,  Germany}
\affiliation{Center for Advanced Systems Understanding (CASUS), D-02826 G\"orlitz, Germany}

\author{Maximilian B\"ohme}

\affiliation{Center for Advanced Systems Understanding (CASUS), D-02826 G\"orlitz, Germany}
\affiliation{Technische  Universit\"at  Dresden,  D-01062  Dresden,  Germany}

\author{Shigenori Tanaka}

\affiliation{Graduate School of System Informatics, Kobe University, Kobe 657-8501, Japan}

\author{Jan Vorberger}

\affiliation{Helmholtz-Zentrum Dresden-Rossendorf (HZDR), D-01328 Dresden, Germany}

\begin{abstract}
We present an \emph{Effective Static Approximation} (ESA) to the local field correction (LFC) of the electron gas that enables highly accurate calculations of electronic properties like the dynamic structure factor $S(q,\omega)$, the static structure factor $S(q)$, and the interaction energy $v$.
{The ESA combines the recent neural-net representation [\textit{J.~Chem.~Phys.}~\textbf{151}, 194104 (2019)] of the temperature dependent LFC in the exact static limit with a consistent large wave-number limit obtained from Quantum Monte-Carlo data of the on-top pair distribution function $g(0)$. It is suited for a straightforward integration into existing codes.}
{
We demonstrate the importance of the LFC for practical applications by re-evaluating the results of the recent {X-ray Thomson scattering experiment on aluminum} by Sperling \textit{et al.}~[\textit{Phys.~Rev.~Lett.}~\textbf{115}, 115001 (2015)]. We find that an accurate incorporation of electronic correlations {in terms of the ESA} leads to a different prediction of the inelastic scattering spectrum than obtained from state-of-the-art models like the Mermin approach or linear-response time-dependent density functional theory.} 
{Furthermore, the ESA scheme is particularly relevant for the development of advanced exchange-correlation functionals in density functional theory.}
\end{abstract}

\maketitle

Warm dense matter (WDM) -- an extreme state of matter characterized by high densities and temperatures -- has emerged as one of the most challenging frontiers of plasma physics and material science~\cite{fortov_review,review,wdm_book}. These conditions occur in many astrophysical objects such as in the interiors of giant planets~\cite{Militzer_2008,militzer1,manuel}, in brown dwarfs~\cite{saumon1,becker}, and in neutron star crusts~\cite{Daligault_2009}. Moreover, they arise in inertial confinement fusion capsules on their pathway towards ignition~\cite{hu_ICF} and are potentially relevant for the understanding of radiation damage in both fission and fusion reactor walls~\cite{reactor}. Furthermore, they apply to the novel field of hot-electron chemistry where the latter are used to accelerate chemical reactions~\cite{Mukherjee2013,Brongersma2015}.

These applications have sparked a surge of activities
in experimental realizations~\cite{falk_wdm} and diagnostics of WDM conditions at intense light sources around the globe, such as at the NIF~\cite{Moses_NIF}, at SLAC~\cite{LCLS_2016}, and at the European X-FEL~\cite{Tschentscher_2017} which have led to experimental breakthroughs over the last years ~\cite{Glenzer_plasmons,ernstorfer2,Fletcher2014,Knudson1455,Kraus2017,Frydrych2020}. While all of these experimental techniques rely on theoretical WDM models to extract observables, an accurate theoretical understanding of WDM is yet missing~\cite{new_POP,wdm_book}.

More specifically, an accurate theoretical description of WDM needs to take into account simultaneously (i) Coulomb coupling effects, (ii) quantum effects, and (iii) thermal excitations. 
In particular, {WDM is characterized by} $r_s\sim\theta\sim1$, where $r_s=\overline{a}/a_\textnormal{B}$ and $\theta=k_\textnormal{B}T/E_\textnormal{F}$ are the usual Wigner-Seitz radius and degeneracy temperature\cite{Ott2018}. 
Under these conditions, thermal density functional theory (DFT)~\cite{KoSh1965,Mermin_1965} has emerged as the work-horse of WDM modeling due to its balance between computational cost and accuracy in terms of an --at least formal-- \textit{ab initio} treatment of the electrons. Despite its current success as a useful technique for the numerical modeling of WDM properties, there are potentially severe limitations for further progress: (1) the accuracy of DFT results crucially depends on an accurate exchange-correlation (XC) functional 
and (2) the computational cost of DFT calculations is too high for on-the-fly diagnostics and interpretation of WDM experiments.

In this regard, the key quantity {for WDM diagnostics is} the dynamic density response function~\cite{kugler1,quantum_theory}
\begin{eqnarray}\label{eq:chi}
\chi\left[G(q,\omega)\right](q,\omega) = \frac{\chi_0(q,\omega)}{1- \frac{4\pi}{q^2}\left(1-G(q,\omega)\right)\chi_0(q,\omega) }\ ,
\end{eqnarray}
where $\chi_0(q,\omega)$ denotes the density response of a non-interacting (ideal) system and the dynamic local field correction (LFC) $G(q,\omega)$ entails {both the frequency and wave number dependence} of XC effects. For example, setting $G(q,\omega)=0$ in Eq.~(\ref{eq:chi}) leads to the well-known random phase approximation (RPA). An accurate knowledge of Eq.~(\ref{eq:chi}) beyond the RPA is paramount for the interpretation of X-ray Thomson scattering (XRTS) experiments~\cite{siegfried_review,kraus_xrts} that presently constitutes the arguably best diagnostics of WDM experiments.

In addition, the LFC is directly proportional to the XC kernel in time-dependent DFT~\cite{dynamic2}, and, moreover, can be used for the construction of an advanced, non-local XC functional for thermal DFT based on the adiabatic-connection formula and the fluctuation-dissipation theorem~\cite{Patrick_JCP_2015,pribram,Goerling_PRB_2019}. 

Recently, Dornheim and co-workers have presented the first accurate {representation of} $G(q,\omega)$ based on \textit{ab inito} path-integral Monte-Carlo (PIMC) data for the warm dense electron gas~\cite{dornheim_dynamic,dynamic_folgepaper,Dornheim_Vorberger_finite_size_2020}. While a full representation of $G(q,\omega)$ covering the entire WDM regime currently remains beyond reach, they have shown that it is often sufficient to replace the dynamic LFC in Eq.~(\ref{eq:chi}) by its static limit, i.e., $G(q)=G(q,0)$. 
It has, indeed, been demonstrated that this \emph{static approximation} $\chi^\textnormal{static}(q,\omega)=\chi[G(q,0)](q,\omega)$ yields highly accurate results for the dynamic structure factor (DSF) $S(q,\omega)$ and related quantities~\cite{hamann_prb_20}. This is a key finding, as $G(q)$ is available as a neural-net representation~\cite{dornheim_ML} that covers the entire relevant range of $r_s$ and $\theta$.

Yet, as we demonstrate in this Letter, the \emph{static approximation} induces a significant bias for medium to large wave numbers $q$, which in turn makes $\chi^\textnormal{static}(q,\omega)$ unsuitable for many applications like the construction of advanced XC functionals for DFT. 
To overcome this severe limitation, we present the \emph{effective static approximation} (ESA) to the LFC given in Eq.~(\ref{eq:ESA}).
It is constructed on the basis of the machine-learning representation of $G(q)$ for small $q$ and, in addition, obeys the consistent asymptotic behaviour in the limit of large wave numbers~\cite{stls}. 
Thus, the ESA yields remarkably accurate results for electronic properties like $S(q,\omega)$, its normalization $S(q)$ [Eq.~(\ref{eq:SSF})], and the interaction energy $v$ [Eq.~(\ref{eq:v})] over the {entire WDM regime without any additional computational} cost compared to the RPA.

\begin{figure}[ht]
\centering
\includegraphics[width=0.368\textwidth]{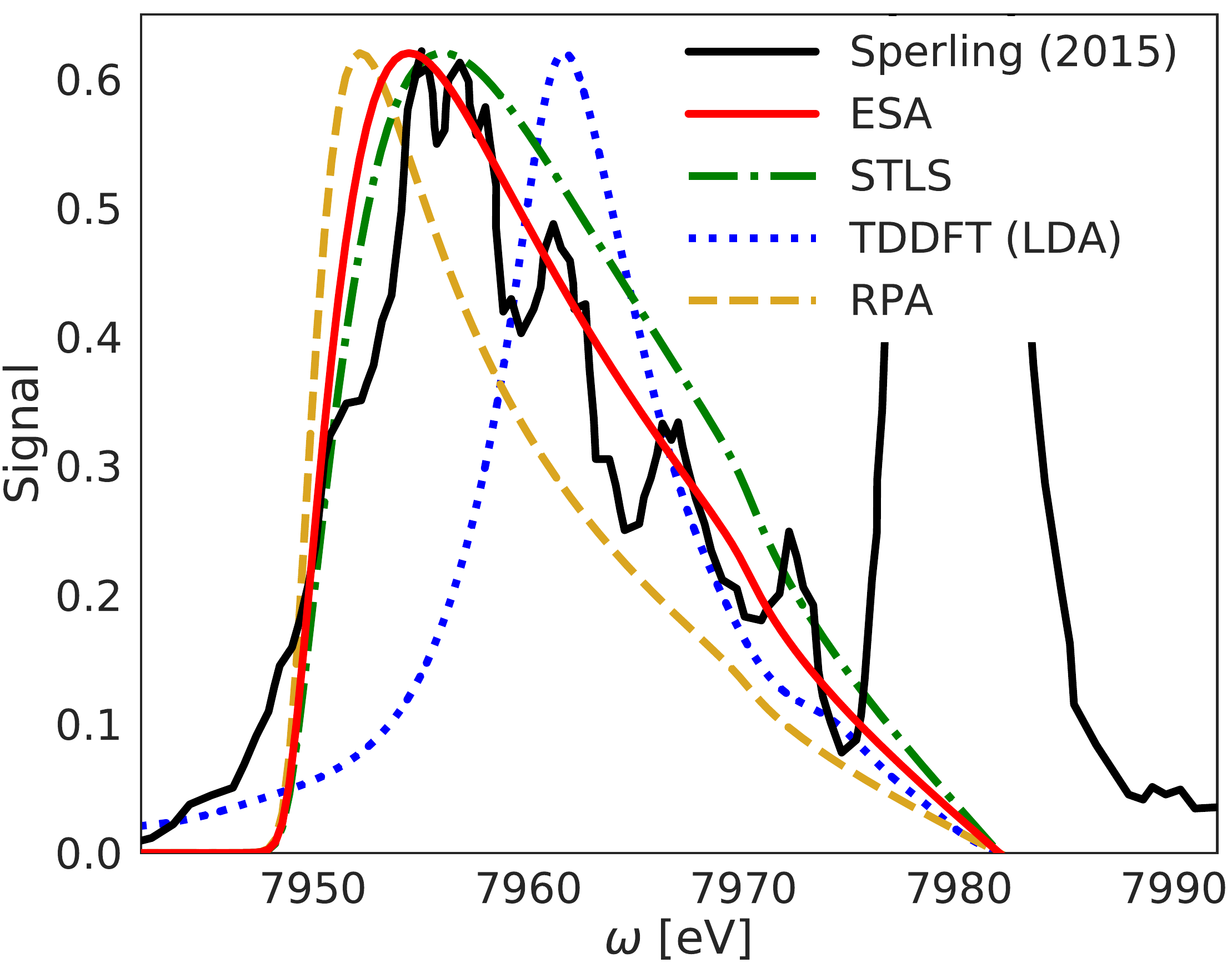}
\caption{The deconvolved XRTS signal in isochorically heated aluminum\cite{SGL15} is compared with the DSF from the ESA (red), STLS (green), time-dependent DFT (blue), and RPA (yellow), all of them computed at a temperature of 0.3~eV. 
}
\label{fig:xrts_aluminum-sperling}
\end{figure}

The ESA is, furthermore, directly applicable as a practical method for the rapid diagnostics of XRTS signals. In Fig.~\ref{fig:xrts_aluminum-sperling}, we demonstrate its utility for the recent XRTS experiment on isochorically heated aluminum by Sperling \textit{et al.}~\cite{SGL15}.
We find a significant improvement over standard dielectric models and a remarkable agreement with the experimental data, even when compared to computationally more complex first-principles techniques such as time-dependent DFT. 

Finally, the proposed ESA enables wide applications beyond XRTS and XC functionals~\cite{Cayzac2017,zhandos_stopping,transfer1,ceperley_potential,zhandos1,zhandos2,new_POP,Diaw2017,zhandos_QHD}.

\textbf{Results.} We begin with benchmarking the \emph{static approximation} against accurate QMC results both for the static structure factor (SSF) $S(q)$ and the interaction energy $v$.
To this end, {we make use of the fluctuation-dissipation theorem~\cite{quantum_theory} 
\begin{eqnarray}\label{eq:FDT}
S(\mathbf{q},\omega) = - \frac{ \textnormal{Im}\chi(\mathbf{q},\omega)  }{ \pi n (1-e^{-\beta\omega})}
\end{eqnarray}
which relates the dynamic density response function $\chi(q,\omega)$ to the DSF $S(q,\omega)$, where $n$ denotes the density and $\beta$ the inverse temperature.}
We note that an extensive analysis of the DSF computed within the \emph{static approximation} has been presented elsewhere~\cite{dornheim_dynamic,dynamic_folgepaper} and need not be repeated here.
The corresponding SSF, defined 
as
\begin{eqnarray}\label{eq:SSF}
S(q) = \int_{-\infty}^\infty \textnormal{d}\omega\ S(q,\omega)\ ,
\end{eqnarray}
\begin{figure}\centering
\hspace*{-0.2cm}\includegraphics[width=0.425\textwidth]{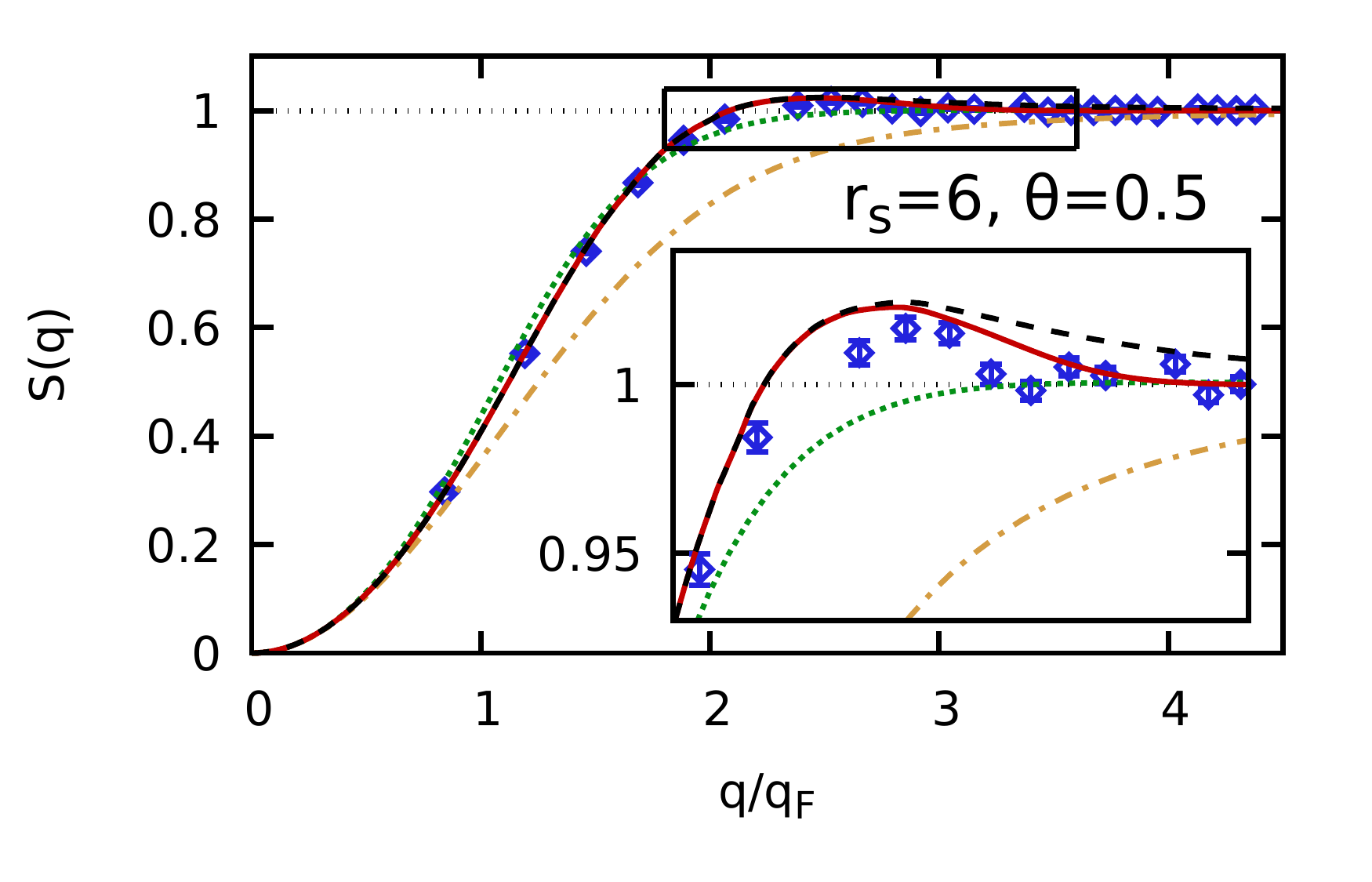}\\\vspace*{-1.22cm}\includegraphics[width=0.415\textwidth]{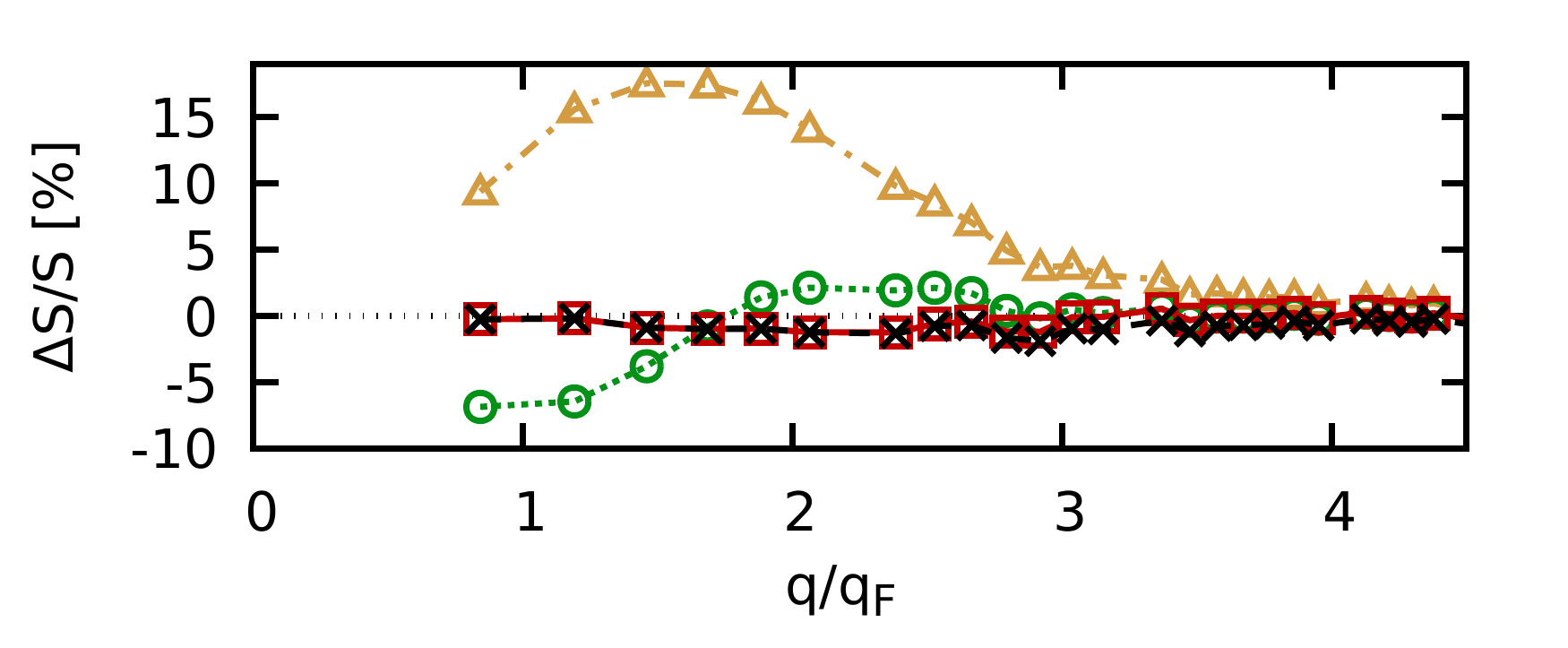}\\\vspace*{-0.34cm}
\caption{\label{fig:Sq}
Top: SSF of the UEG at $r_s=6$ and $\theta=0.5$. Blue diamonds: PIMC, solid red: ESA; dashed black: \emph{static approximation}; dotted green: STLS~\cite{stls,stls2}; dash-dotted yellow: RPA. Bottom: Relative deviation {from the PIMC results}.
}
\end{figure}  
is shown in the top panel of Fig.~\ref{fig:Sq} {for the conditions $r_s=6$ and $\theta=0.5$ which are realized experimentally in hydrogen jets~\cite{Zastrau} and evaporation experiments~\cite{benage,karasiev_importance,low_density1,low_density2}. 
Due to the pronounced impact of electronic XC effects~\cite{low_density1}, these conditions are challenging from a theoretical perspective and are, therefore, well-suited to benchmark different models.}
The blue diamonds correspond to PIMC data and are exact within the given error bars. The dashed black line is obtained from the \emph{static approximation} where the exact static limit of $G(q,\omega)$ available as a neural-net representation~\cite{dornheim_ML} was used as input. Remarkably, it is in striking agreement with the exact PIMC results with a maximum deviation of $\sim 1\%$ (see the bottom panel). As a reference, we also include the SSF computed within the RPA (dash-dotted yellow) and the LFC of Singwi \textit{et al.}~\cite{stls_original,stls,stls2} (STLS, dotted green). As one might expect, the RPA gives a poor description at these conditions, reflected by the relative deviation exceeding $15\%$. The STLS formalism is based on an approximate closure relation for $G(q,0)$ and leads to a substantial improvement over RPA. Nevertheless, there are still systematic errors: the relative deviation is about $8\%$ and the correlation-induced maximum in $S(q)$ that appears at $q\approx2.2q_\textnormal{F}$ is not reproduced by STLS. We thus conclude that the \emph{static approximation} provides a highly accurate description with negligible computational cost even at such challenging conditions; more examples can be found in the Supplemental Material~\cite{supplement}.

\begin{figure}\centering
\includegraphics[width=0.415\textwidth]{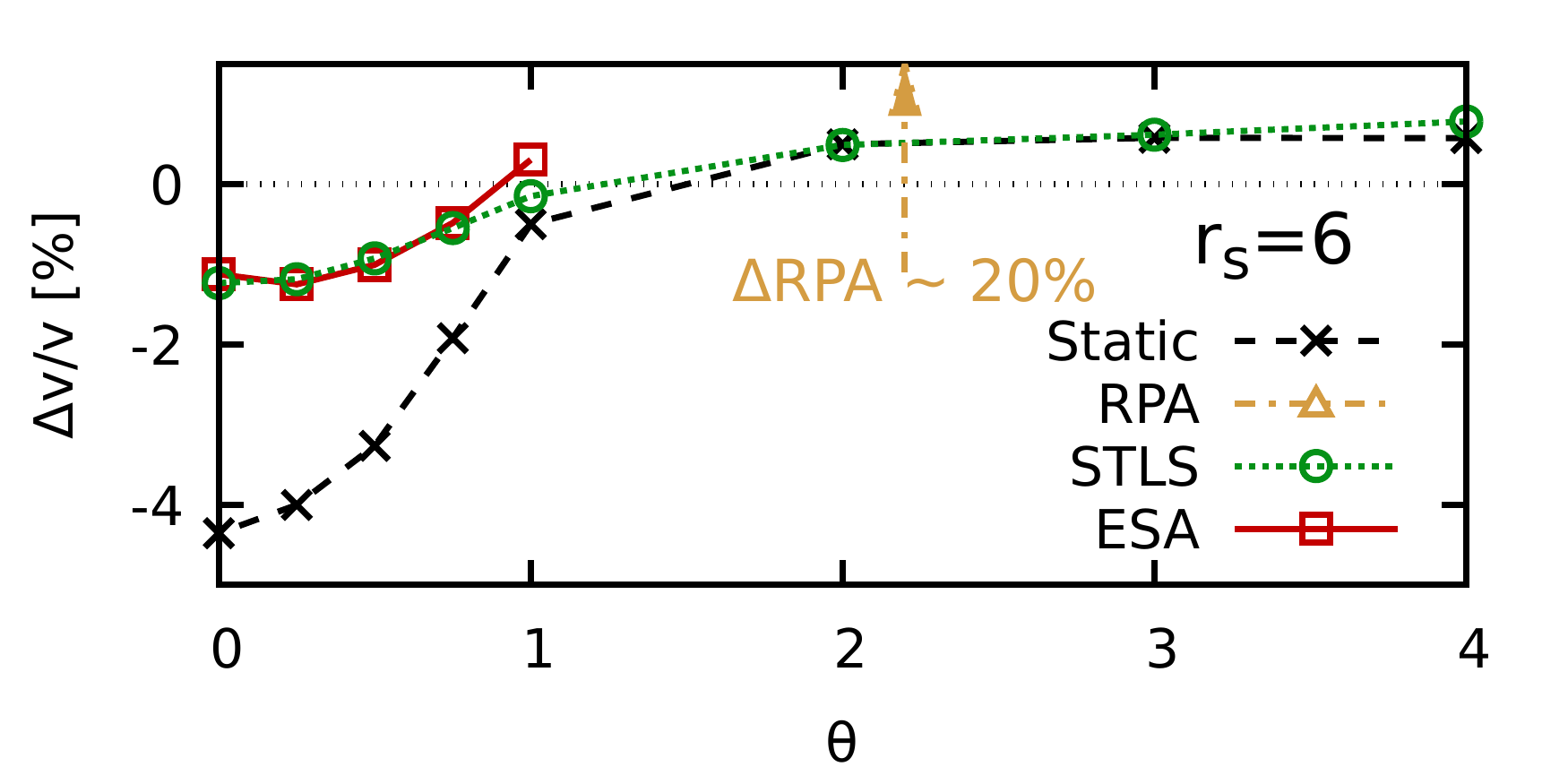}\\\vspace*{-0.91cm}
\hspace*{-0.3995cm}\includegraphics[width=0.44\textwidth]{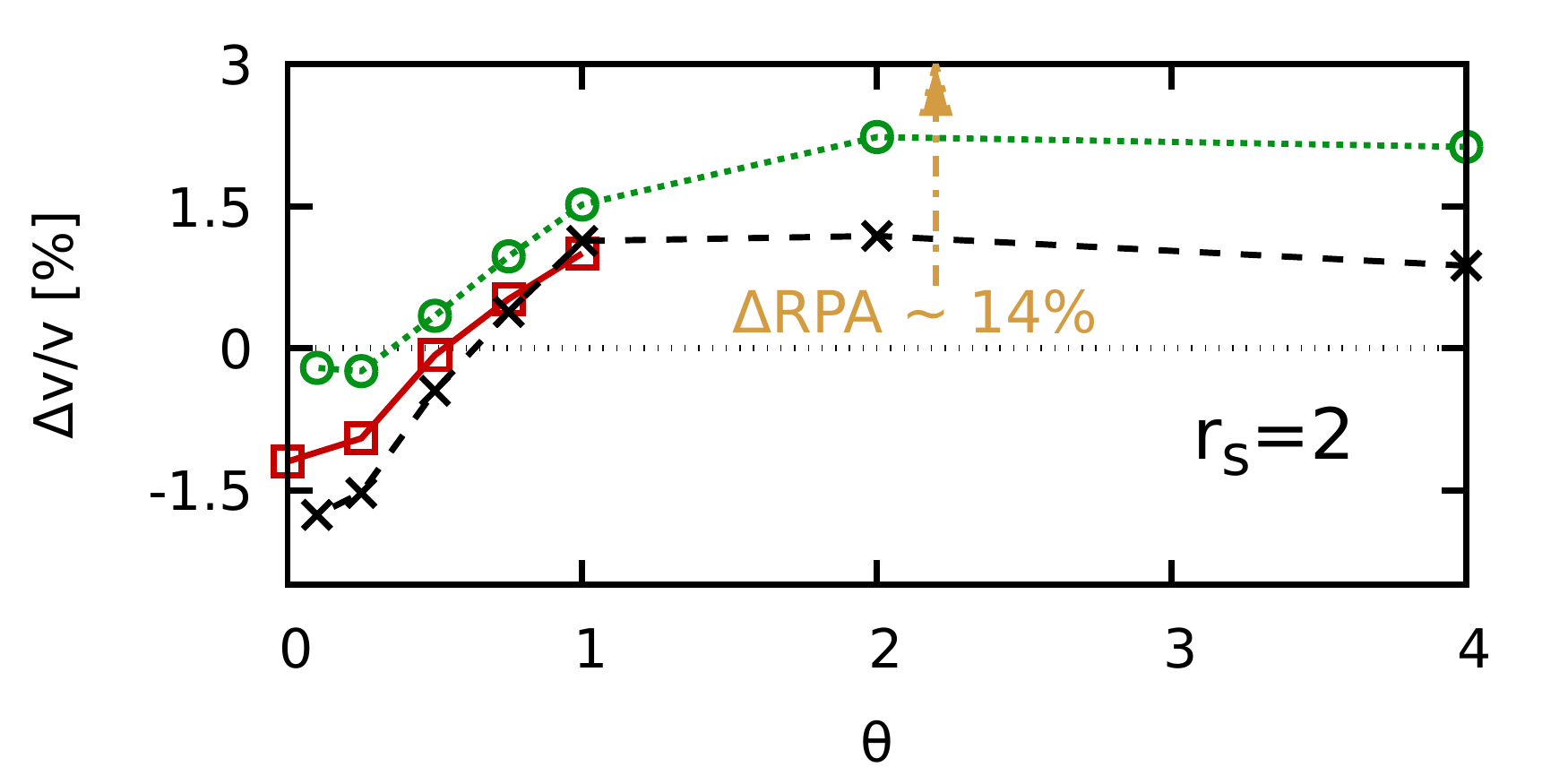}
\caption{\label{fig:v}
Relative difference in the interaction energy per particle $v$ as compared to the accurate parametrization of the exchange--correlation free energy $f_\textnormal{xc}$ by Groth \textit{et al.}~\cite{groth_prl}.
}
\end{figure}

Let us for now postpone the discussion of the ESA (solid red) in Fig.~\ref{fig:Sq}, and investigate the interaction energy of the UEG, computed from $S(q)$ via~\cite{review}
\begin{eqnarray}\label{eq:v}
v = \frac{1}{\pi} \int_0^\infty \textnormal{d}q\ \left[
S(q)-1
\right]\ .
\end{eqnarray}
{In Fig.~\ref{fig:v} we illustrate the relative accuracy of $v$ within different theories over the relevant $\theta$ range and at two relevant values of the density parameter $r_s$. The reference result is the QMC-based parametrization by Groth \textit{et al.}~\cite{groth_prl}, which is exact to within $\sim0.3\%$.}
The top panel corresponds to $r_s=6$, which is most challenging for most theories due to the strong coupling strength. Unsurprisingly, RPA is highly inaccurate over the entire $\theta$ range with a relative deviation of $\sim 20\%$, whereas STLS and the \emph{static approximation} exhibit some interesting behavior: For $\theta\gtrsim1$, both STLS and the \emph{static approximation} are basically exact and can hardly be distinguished from each other. For $\theta < 1$, the STLS curve does still not exceed deviations of $2\%$, whereas the quality of the \emph{static approximation} deteriorates as $\theta$ decreases with a systematic deviation of almost $5\%$ in the ground state.
Let us first consider the comparably high accuracy of STLS for $v$. Evidently, this is not due to an inherently correct physical description of the system, as STLS does not reproduce important trends (see Fig.~\ref{fig:Sq}). The high accuracy for $v$ is rather the result of a fortunate cancellation of errors in $S(q)$ when inserted into Eq.~(\ref{eq:v}), as it is too large in the small- and too low in the high-wave number regime. In contrast, the \emph{static approximation} provides a high-quality description of $S(q)$ for all $q$,
but converges too slowly towards unity for large $q$ (see the inset in Fig.~\ref{fig:Sq}). While this bias is relatively small for each individual $q$ value, the corresponding error in $v$ accumulates under the integral in Eq.~(\ref{eq:v}) and leads to a substantial bias in the interaction energy.

To develop an improved theory based on the \emph{static approximation} without this obstacle, we have to first understand its origin. {Our analysis centers on the well-known asymptotic behaviour of static LFCs}~\cite{stls} for large $q$
\begin{eqnarray}\label{eq:limit}
\lim_{q\to\infty} G(q) = 1-g(0)\ ,
\end{eqnarray}
where $g(0)$ is the \emph{on-top} pair distribution function (PDF), i.e., the PDF at zero distance. 

\begin{figure}\centering
\includegraphics[width=0.415\textwidth]{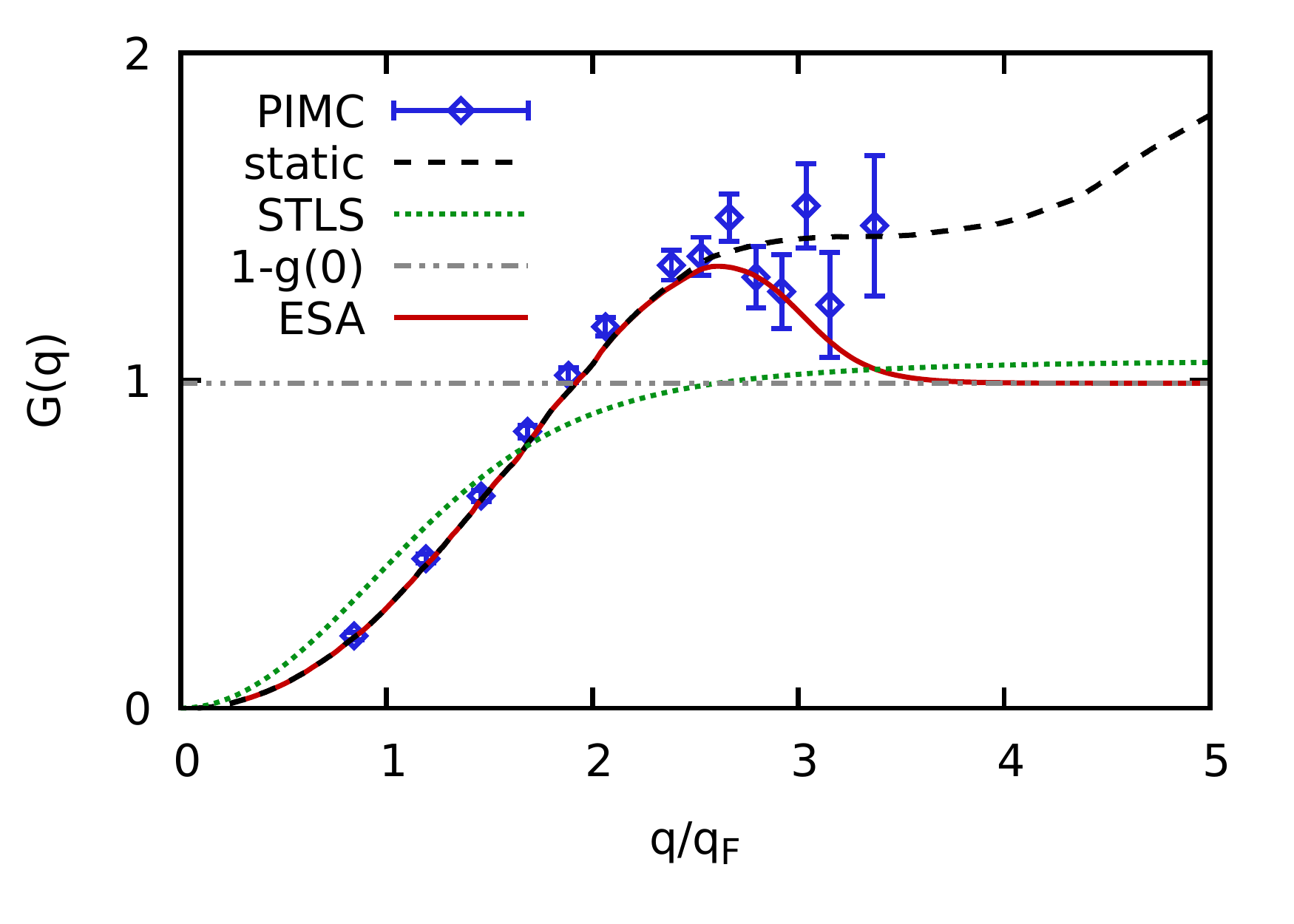}\\\vspace*{-0.36cm}
\caption{\label{fig:Gq}
Wave-number dependence of the local field correction $G(q)$ at $r_s=6$ and $\theta=0.5$. The \emph{static} curve has been obtained from the neural-net given in Ref~\cite{dornheim_ML}, and the ESA curve corresponds to Eq.~(\ref{eq:ESA}). The parametrization of the on-top PDF $g(0)$ is given in the Supplemental Material~\cite{supplement}.
}
\end{figure}

This is illustrated in Fig.~\ref{fig:Gq}, where we show $G(q,0)$ again at $r_s=6$ and $\theta=0.5$. The dotted green curve corresponds to STLS, which is an example for such a static theory obeying Eq.~(\ref{eq:limit}), i.e., it converges towards a constant for large $q$. As a side note, we mention that $G_\textnormal{STLS}(q\to\infty)>1$, which leads to an unphysical \emph{negative} value for $g(0)$, see also Refs.~\cite{quantum_theory,Kumar_PRB_2009}.
The blue diamonds in Fig.~\ref{fig:Gq} have been obtained from a PIMC simulation (see Refs.~\cite{dornheim_ML,dornheim_HEDP,dornheim_electron_liquid} for details) and are exact within the given error bars. 
The increasing level of noise towards large $q$ is due to the reduced impact of $G(q,\omega)$ [see Eq.~(\ref{eq:chi})], which is further exacerbated by the fermion sign problem~\cite{dornheim_sign_problem,dornheim_permutation_cycles,troyer}.
{ 
Similarly as in Fig.~\ref{fig:Sq}, we find that STLS does not give a qualitatively correct description of the $q$-dependence, and, in addition, also violates the compressibility sum rule for small $q$, see Ref.~\cite{stls2}. 
The dashed black line depicts the neural-net representation of the exact, \emph{static} LFC from Ref.~\cite{dornheim_ML} and it is in excellent agreement with the PIMC data. We note that the PIMC data were not used as input for the neural net and, thus, constitute a valuable validation of the dashed black curve for $q\lesssim3q_\textnormal{F}$, whereas the PIMC error bars are too large for larger $q$ to assess its quality.
Further, the black curve increases monotonically with $q$ and, thereby, violates Eq.~(\ref{eq:limit}).
}

In fact, it can be shown that this long-wave number behavior of the exact $G(q,0)$ is responsible for the unphysically slow convergence of $S(q)$ towards unity within the \emph{static approximation}~\cite{supplement}. Methods like STLS~\cite{stls_original,stls2,stls} and other static dielectric theories~\cite{tanaka_hnc,stolzmann} are based on a LFC independent of $\omega$, but still coupled to $S(q)$ via some form of closure relation. Therefore, these theories do not necessarily constitute an approximation to $\lim_{\omega\to0}G(q,\omega)$, but can be viewed as a frequency-averaged LFC, i.e., an LFC that is meaningful for quantities that involve a frequency-integral like $S(q)$ or $v$. In contrast, the \emph{static approximation} is based on the exact $\omega\to0$ limit of $G(q,\omega)$, which gives remarkably high-quality results for $S(q,\omega)$ and $S(q)$, but induces small, yet significant, unphysical effects that accumulate under a wave-number integral. In addition to the bias in $v$, the slow convergence of $S(q)$ also induces a divergent on-top PDF~\cite{supplement}, and, thus, substantially limits the usefulness of the \emph{static approximation} that is directly based on the \emph{static} LFC from Ref.~\cite{dornheim_ML}.

{To overcome these limitations, we introduce the \emph{effective static approximation} (ESA) as the central result of this paper.} It combines the exact $G(q,0)$ for $q\lesssim3q_\textnormal{F}$ with the appropriate long-wave number limit in Eq.~(\ref{eq:limit}), thereby, ensuring a proper convergence of $S(q)$ and the correct on-top PDF $g(0)$. The resulting LFC has the form
\begin{eqnarray}\label{eq:ESA}
G_\textnormal{ESA}(q) = A(q) \left(1-g(0)\right)+ G_\textnormal{nn}(q) \left(
1-A(q)
\right)\ ,
\end{eqnarray}
where $A(q)$ is a simple activation function~\cite{supplement} and $G_\textnormal{nn}(q)$ corresponds to the neural-net from Ref.~\cite{dornheim_ML}.
We note that the specific form of the activation function is not particularly important for the ESA as long as the conditions $A(0)=0$ and $A(q\to\infty)=1$ are satisfied; the empirical choice for $A(q)$ used in this work is discussed in the Supplemental Material~\cite{supplement}.
In addition, we have constructed an analytical parametrization of $g(0)$ that combines the ground-state results by Spink \textit{et al.}~\cite{Spink_Drummond_PRB_2013} with the restricted PIMC results by Brown \textit{et al.}~\cite{Brown_ethan} at finite $\theta$. Both the functional form and the corresponding fit parameters are given in the Supplemental Material~\cite{supplement} and can be used for other applications~\cite{STAROSTIN200064,Coraddu2006,Sjostrom_Daligault_PRB}.

The resulting LFC is shown as the red curve in Fig.~\ref{fig:Gq}
and does indeed smoothly combine the exact $G(q,0)$ with the consistent limit in Eq.~(\ref{eq:limit}). The impact of this improvement is illustrated in Fig.~\ref{fig:Sq}, where the ESA reproduces the accurate $S(q)$ from the \emph{static approximation} for $q\lesssim3q_\textnormal{F}$, but, in addition, exhibits a much faster convergence to unity for large $q$. As expected, this leads to substantially improved results for integrated quantities, such as interaction energies with an accuracy of $\sim1\%$ (Fig.~\ref{fig:v}). The improved results for $v$ are also shown at a higher density, $r_s=2$, in the bottom panel of the same figure. Additional results of $S(q)$ and $G(q)$ are shown in the Supplemental Material~\cite{supplement}.

Up to this point, we have shown that the proposed ESA is capable of yielding highly accurate results for $S(q,\omega)$, $S(q)$, and $v$ without any additional computational cost compared to the RPA.

We conclude this Letter by turning to an actual application of the ESA. We demonstrate its utility as a first-principles method for the rapid interpretation of XRTS signals.  
Specifically, we consider the XRTS experiment on isochorically heated aluminum by Sperling \textit{et al.}~\cite{SGL15} shown in Fig.~\ref{fig:xrts_aluminum-sperling} and demonstrate the impact of electronic XC effects included in the ESA. We compare the deconvolved scattering signal collected from the corresponding XRTS experiment at a scattering angle of $\theta=24 ^{\circ}$ (black)  with several theoretical predictions of the DSF.
The theoretical predictions are renormalized with respect to the peak at around 7958 eV in the experimental data. 
The ESA (red) is in remarkable agreement with the experimental data, while coming at a computational cost of the simple RPA. 
The RPA (yellow) yields only qualitative agreement. 
While the static LFC within STLS (green) is closer to the ESA result, it also comes at a higher computational cost compared to the ESA, because a self-consistent set of equations for the static structure factor, the dielectric function, and the static LFC needs to be solved.
The computationally more complex time-dependent DFT (blue) within the adiabatic LDA also yields only qualitative agreement (see also Ref.~\cite{Mo2018}) and is, furthermore, orders of magnitudes more expensive than the ESA. 
The results shown here are computed at a temperature of $0.3$ eV. Additional details and results at the nominal temperature of 6.0~eV are given in the Supplemental Material~\cite{supplement}.
Furthermore, in contrast to common, low-cost dielectric models based on phenomenological parameters\cite{HuPe1957,Mermin_1965,stls_original,Fortmann2010,WiGaGaLe2017,WiSpFrRe2018}, the ESA provides a consistent prediction of XRTS signals from first principles~\cite{supplement} and does not rely on any phenomenological parameters.   

\textbf{Discussion.} In summary, we have presented the ESA which is capable of providing highly accurate results for electronic properties like $S(q,\omega)$, $S(q)$, and $v$, without any additional computational cost compared to standard RPA calculations. 
{
We expect the ESA to replace all known RPA+LFC combinations. 
The ESA is likely to have tremendous impact in} a large number of applications beyond the interpretation of XRTS experiments, such as in the calculation of stopping powers~\cite{Cayzac2017,zhandos_stopping}, energy relaxation rates~\cite{transfer1}, and electrical or thermal conductivities~\cite{Reinholz_PRE_2000}. Other examples include
the construction of effective potentials~\cite{ceperley_potential,zhandos1,zhandos2}, quantum hydrodynamics~\cite{new_POP,Diaw2017,zhandos_QHD}, and modeling of high energy density physics phenomena with average-atom codes\cite{SHWI2007:equation, PhysRevE.89.023108}. 
Finally, we point out that the ESA is particularly relevant for wave-number averaged quantities like $v$, which is of key importance for constructing advanced XC functionals~\cite{Patrick_JCP_2015,pribram,Goerling_PRB_2019}. 
A python-based implementation of the ESA is freely available online~\cite{supplement} and can be easily incorporated into existing codes. Moreover, the ESA will be included in a novel open-source XRTS code that is being developed jointly by HZDR and CASUS.

\begin{acknowledgments}
\section*{Acknowledgments}
We acknowledge helpful feedback by M.~Bonitz and M.~Bussmann.

This work was partly funded by the Center for Advanced Systems Understanding (CASUS) which is financed by the German Federal Ministry of Education and Research (BMBF) and by the Saxon Ministry for Science, Art, and Tourism (SMWK) with tax funds on the basis of the budget approved by the Saxon State Parliament.
Part of AC's initial work on this was performed under the Laboratory Directed Research and Development program at Sandia National Laboratories (SNL). SNL is managed and operated by NTESS under DOE NNSA contract DE-NA0003525.
We gratefully acknowledge computation time on the Bull Cluster at the Center for Information Services and High Performace Computing (ZIH) at Technische Universit\"at Dresden, and at the Norddeutscher Verbund f\"ur Hoch- und H\"ochstleistungsrechnen (HLRN) under grants shp00015 and shp00026.
\end{acknowledgments}
\bibliography{bibliography}{}

\end{document}